\documentclass[preprint,12pt]{elsarticle}

\usepackage{amssymb}
\usepackage{amsmath}
\usepackage{xcolor}

\journal{European Journal of Control}

\begin{document}

\begin{frontmatter}



\title{State estimation for a class of nonlinear time-varying uncertain system under multiharmonic disturbance}
\tnotetext[t1]{The study was supported by the Ministry of Science and Higher Education of the Russian Federation, state assignment No. 2019-0898.}


\affiliation[inst1]{organization={Control Systems and Robotics Department, ITMO University},
            addressline={Kronverksky Pr. 49, bldg. A}, 
            city={St. Petersburg},
            postcode={197101}, 
            country={Russia}}

\affiliation[inst3]{organization={Inria,  University Lille, CNRS, UMR 9189 - CRIStAL},
            addressline={F-59000}, 
            city={Lille},
            country={France}}

\author[inst1]{Alexey A. Margun}
\author[inst1]{Van H. Bui}
\author[inst1]{Alexey A. Bobtsov}
\author[inst3]{Denis V. Efimov}

\begin{abstract}
The paper considers the observer synthesis for nonlinear, time-varying plants with uncertain parameters under multiharmonic disturbance. It is assumed that the relative degree of the plant is known, the regressor linearly depends on the state vector and may have a nonlinear relationship with the output signal. The proposed solution consists of three steps. Initially, an unknown input state observer is synthesized. This observer, however, necessitates the measurement of output derivatives equal to the plant's relative degree. To relax this limitation, an alternative representation of the observer is introduced. Further, based on this observer,  the unknown parameters and disturbances are reconstructed using an autoregression model and the dynamic regressor extension and mixing (DREM) approach. This approach allows the estimates to be obtained in a finite time.  Finally, based on these estimates, an observer has been constructed that does not require measurements of the output derivatives. The effectiveness and efficiency of this solution are demonstrated through a computer simulation.
\end{abstract}



\begin{keyword}
unknown input observer \sep disturbance \sep time-varying plant \sep nonlinear system
\end{keyword}

\end{frontmatter}


\section{Introduction}
\label{sec:intro}
To enhance the effectiveness of automatic control systems, it is crucial to implement state feedback, necessitating the measurement of state vector variables. However, achieving this in real-world applications poses significant challenges. The installation of sensors for measuring all necessary variables can be prohibitively expensive and sometimes technologically unfeasible. To address this issue, control theory has introduced the concept of observers. These tools provide the estimation of state variables without direct measurement, offering a practical solution to the limitations posed by sensor implementation  (see, for example, \cite{kailath1980linear,2,3,4}).

In 1964, Luenberger pioneered the theory of state observer synthesis for linear systems \cite{4}. However, as science and technology have advanced, linear dynamic models with constant parameters often fail to meet practical demands. Real-world systems are frequently affected by various unknown external perturbations, presenting challenges in developing algorithms for constructing state observers for nonlinear, time-varying systems under such disturbances. To tackle these challenges, several classic methods have been proposed.

One of the most popular approaches consists in (piecewise) approximation of nonlinearities by linear functions \cite{linearization5,linearization7,linearization9,approximation6,approximate8}. This method, illustrated in various studies, includes the use of exponential functions \cite{approximation6}, linear-quadratic objective function minimization \cite{linearization7}, and Taylor series expansion combined with the least squares method, utilizing Moore-Penrose pseudo-inverse transformation \cite{approximate8}. This approach offers ease in parameter selection and practical application for specific technical systems. It is notably effective where nonlinear filters are impractical or unfeasible. However, it faces challenges and limitations in systems with complex mathematical models or when precise perturbation distributions are required, as inaccuracies in linearization can sometimes result in system instability.

Methods employing the extended Kalman filter \cite{Kalman13,kalman14,kalman15} are popular in various practical applications. The key advantage of this approach is its suitability for nonlinear systems and its capacity to estimate both the system state and the parameters of disturbances, described by probabilistic models. However, this method also requires model linearization and knowledge of perturbation distributions, whose absence may lead to estimation errors.

In \cite{sliding16,sliding17}, a sliding mode-based observer synthesis method for state vector estimation is introduced. The primary advantage of this method is its robustness to parametric disturbances, owing to its invariance properties. However, the occurrence of sliding modes introduces challenges, notably oscillations and high-frequency switching in the control channel, which are significant concerns in the application of this technique.

An alternative approach involves transforming the original system into a linear regression form with an unknown parameter vector \cite{5_bobtsov2022adaptive,6_bobtsov2021state}. This transformation allows the application of various identification techniques, such as the least squares method \cite{35lyung1991identification}, gradient descent \cite{35lyung1991identification}, dynamic regressor extension and mixing method \cite{36vediakova2020drem,37ovcharov2020finite,38ortega2020new}, etc. The primary advantage of this approach is the independence of the identification block from the control law, allowing the use of diverse control methods without altering the identification algorithm. However, a limitation of such identification techniques is the necessity to fulfill the persistent excitation condition for the regressor. Additionally, there is no universally applicable method for different plant classes; each system requires a unique parameterization solution. A recent development in this area focuses on constructing observers for uncertain plants with output-dependent unknown parameters \cite{2pyrkin2023adaptive}.

A specific class of solutions in the field of observer synthesis provides a state estimation in the presence of unmeasured input signals. This class of solutions is called Unknown Input Observers (UIOs) \cite{40hou1992design,41chen1996design,42warrad2018design}. These algorithms ensure effective functioning even in the presence of disturbances and other factors that affect the signals in the control channel. They are designed to estimate state vector despite the presence of external influencing factors. Unknown input observers were first introduced by N. Kobayashi \cite{1982observer} in 1982. This result has been further developed for discrete-time systems \cite{50warrad2018design,51sharma2016unknown}, various nonlinear systems \cite{50warrad2018design,51sharma2016unknown,57coutinho2022sufficient}, and time-varying systems \cite{57coutinho2022sufficient,60aguilera2012sensor}.

For time-varying systems a common approach is to transform the original plant into a polytopic form with weight functions that depend on parameters. To construct the observer and ensure its stability, Lyapunov function framework and  solution of a system of matrix inequalities are typically used. For example, in \cite{56gomez2019sensor} an unknown input observer was developed for Takagi-Sugeno continuous and discrete-time systems which are analogous to polytopic nonstationary systems. Further, this method was extended to a class of nonlinear systems in \cite{57coutinho2022sufficient}.

Generally, constructing unknown input observers is feasible primarily for systems where the relative degree between unknown inputs and the output signal is one. For plants with a higher relative degree, stringent conditions must be met to build an observer. For instance, researches like \cite{62ichalal2015unknown,63floquet2007sliding} necessitate measuring the derivatives of the system’s output signal. In works such as \cite{57coutinho2022sufficient,58wang2021novel}, the system's original dynamic equation must be separable to isolate the observable components in the output.

In recent studies various methods have been developed for estimating state vectors in nonlinear systems. However, these solutions face significant constraints. For example, paper \cite{2pyrkin2023adaptive} presents a state vector estimation method for nonlinear systems with unknown parameters, where the regressor is dependent on the output. The dependence of regressor on unknown state vectors adds complexity to the problem. Research \cite{farza2009adaptive} addresses this challenge. However, proposed solution is applicable only to a class of stationary systems, and it is difficult to regulate the estimation error convergence rates.

In this paper, we combine two methodologies to develop a novel, comprehensive solution for nonlinear time-varying systems based on the unknown input and the identification-based observer synthesis. This integrated approach aims to address the limitations inherent in each individual method, facilitating efficient operation across a broader spectrum of systems providing the estimation of the state, with the reconstruction of the parameters and the unknown input in a finite time.

The structure of the paper is outlined as follows: Section 2 presents the problem statement. Section 3 proposes the unknown input observer synthesis algorithm, first, assuming the availability of the derivatives of the measured output signal, and second, relaxing this constraint. Section 4 introduces an algorithm to estimate the unknown parameters and external harmonic disturbances, and state observer based on these estimates that does not require output derivatives. Finally, the paper concludes with simulation results that demonstrate the efficiency of the proposed method.

\section{Problem Statement}

Consider a class of single input single output plants described by equations
\begin{equation}
    \label{plant}
    \left\{ 
        \begin{aligned}
            &\dot{x}(t) = Ax(t)+B[u(t)+\varphi^T(x(t),y(t),t)\theta(t)+f(t)],\\
            &y(t)=Cx(t),
        \end{aligned}
    \right.
\end{equation}
where $x(t)\in\mathbb{R}^n$ is an unmeasured state vector, $u(t)\in\mathbb{R}$ is a known input signal, $\varphi(x,y,t)$ is a known vector function, $\theta(t) \in \mathbb{R}^w$ is an unknown vector of time-varying parameters, $y(t)$ is a measured output signal, $f(t) \in \mathbb{R}$ is an unknown multiharmonic disturbance; $A$, $B$, $C$ are known matrices with appropriate dimensions; plant has relative degree between the input and output $r\le n$ that means $CB=CA^{j}B=0$ for $j=1,\dots,r-2$. The system under consideration is time-varying, containing nonlinear functions of the output signal and the product of unknown variable parameters and an unmeasured state vector. The last factor does not allow the use of such well-known solutions as the extended Kalman filter, the internal model principle, etc. This class of models describes the dynamics of such technical systems as the bicycles \cite{bicycle_model}, surface ships \cite{ship_model}  and many others.

We will need the following assumptions:
\begin{enumerate}
    \item The external disturbance $f(t)$ is represented as $f(t) = \sum\limits_{i = 1}^q {{R_i}\sin ({\omega _i}t + {\phi _i})}$, where $R_i$, $\omega_i$, $\phi_i$ are unknown amplitudes, frequencies and phases, respectively, $q$ is a known number of harmonics.
    \item Pair $(A, C) $ is observable.
    \item Matrix $B$ has full column rank and matrix $C$ has full row rank.
    \item Time-varying parameters vector $\theta(t)$ is an output of a linear generator:
    \begin{gather*}
        \theta(t) = H\xi(t),\\
        \dot{\xi}(t)=\Gamma\xi(t),
    \end{gather*}
    where $\xi(t)\in\mathbb{R}^m$ is a generator state vector with unknown initial conditions $\xi(0)$. Constant matrices $H$ and $\Gamma$ are known.
    \item The regressor is a linear function of the state weighted by a continuous nonlinear function of the output and the time:
    \begin{gather*}
        \varphi(x,y,t) = \alpha(y,t) x,
    \end{gather*}
    where $\alpha:\mathbb{R}^{2}\to\mathbb{R}$ is known.
\end{enumerate}

The goal is synthesis of an adaptive observer for the plant (\ref{plant}) under the introduced restrictions, which guarantees realization of the following relations:
$$  {\lim_{t\to+\infty}|\hat x(t)-x(t)|=0} $$
and $\forall t \ge \tau$
$$
{\hat f(t) = f(t)}, {\hat \theta(t) = \theta(t)},
$$
where $\hat{x}(t)$ denotes the estimated state vector, $\hat{f}(t)$  represents the estimated external disturbance, $\hat \theta(t)$ is an estimate for $\theta(t)$, and $\tau > 0$ is a finite time of estimation.

\textbf{Remark 1:} For simplification of the mathematical representation and for the reader's convenience, we assume $u(t) \equiv 0$ in the sequel. 

\section{Unknown Input Observer}
In this section observer synthesis algorithm is given. First, an exponentially converging state UIO is introduced that describes the basic structure of the estimator. Next, an alternative realization of this observer is derived that reduces the number of needed derivatives to $r-1$ and that provides a finite-time estimation of the state.

Consider plant (\ref{plant}) with relative degree $r \le n$. The unknown input observer is formulated as
\begin{equation}
    \label{observer1}
    \dot{\hat{x}}(t) = F\hat{x}(t)+Ly(t)+Gy^{(r)}(t),
\end{equation}
where matrices $F, L$ and $G$ satisfy the following conditions:
\begin{equation}
\label{condition1}
    B-GCA^{r-1}B=0,
\end{equation}
\begin{equation}
\label{condition2}
    M=A-GCA^r,
\end{equation}
\begin{equation}
\label{condition3}
    F=M-LC.
\end{equation}

To demonstrate that (\ref{observer1}) is an observer, consider the error $\tilde{x}(t)=x(t)-\hat{x}(t)$. By differentiating and incorporating (\ref{plant}) and (\ref{observer1}), the dynamic model of $\tilde{x}(t)$ is derived as follows:
$$
    \begin{aligned}
        &\dot{\tilde{x}}(t) =Ax(t)+B[\varphi^T(x,y,t)\theta(t)+f(t)]-F\hat{x}(t)-Ly(t)-Gy^{(r)}(t)=\\
        & = Ax(t)+B[\varphi^T(x,y,t)\theta(t)+f(t)]-F\hat{x}(t) -Ly(t)-GCA^rx(t)-\\
        & -GCA^{r-1}B[\varphi^T(x,y,t)\theta(t)+f(t)].
    \end{aligned}
$$
Applying (\ref{condition2}) and (\ref{condition3}), we get:
\begin{equation}
\label{eq5}
    \dot{\tilde{x}}(t)=F\tilde{x}(t)+(B-GCA^{r-1}B)(\varphi^T(x,y,t)\theta(t)+f(t)).
\end{equation}
Substituting (\ref{condition1}) into (\ref{eq5}) results in:
\begin{equation}
\label{closed}
    \dot{\tilde{x}}(t)=F\tilde{x}(t).
\end{equation}
Obviously, if the matrix $F$ is Hurwitz, then $\tilde{x}(t)$ exponentially converges to zero. 

Constructing the observer as defined in (\ref{observer1}) necessitates solving the system of equations given by (\ref{condition1}), (\ref{condition2}) and (\ref{condition3}). This system is solvable under the following condition \cite{41chen1996design}:
\begin{itemize}
    \item $\textrm{rank}(CA^{r-1}B) = \textrm{rank}(B),$
    \item $(C,M)$ is a detectable pair.
\end{itemize}

Introduce the following auxiliary variables to relax the problem of utilization of the
unmeasured output derivatives (hereinafter omit the time dependence for brevity):
\begin{equation}
\label{auxiliaxry_vars}
\begin{split}
    &z_1 = \hat{x}-Gy^{(r-1)},\\
    &\dot{z}_1 = F\hat{x}+Ly+Gy^{(r)}-Gy^{(r)}=F(z_1+Gy^{(r-1)})+Ly,\\ 
    &z_2 = z_1-FGy^{(r-2)}, \\
    &\dot{z}_2 = F(z_1+Gy^{(r-1)})+Ly-FGy^{(r-1)}=F(z_2+FGy^{(r-2)})+Ly,\\
    \vdots \\
    & z_r = z_{r-1} - F^{r-1}Gy,\\
    & \dot{z}_r = F(z_r+F^{r-1}Gy)+Ly.
\end{split}
\end{equation}

Therefore, the state vector estimate is given by
\begin{equation}
\label{x_estmate}
    \hat{x}=z_1+Gy^{(r-1)} = z_r + F^{r-1}Gy+\ldots +Gy^{(r-1)}.
\end{equation}

If the matrix $L$ is chosen to satisfy $F^r G + L = 0$, then $\dot{z}_r(t) = Fz_r(t)$. In this case, $z_r(t)$  is an exponentially decaying function and state vector estimate takes the form
\begin{equation}
\label{x_estmate_no_z}
    \hat{x}=F^{r-1}Gy+\ldots +Gy^{(r-1)}+e^{Ft}z_r(0).
\end{equation}

\textbf{Theorem 1.} Let $r-1$ derivatives of the output signal $y$ are available for measurement and the matrices $M, L, G$ are chosen to satisfy conditions (\ref{condition1}) -- (\ref{condition3}) with $F$ being Hurwitz, then the observer (\ref{x_estmate}) with the auxiliary variables (\ref{auxiliaxry_vars}) provides estimation of the state vector $x(t)$ with exponential convergence. If additionally 
\begin{equation}
L + F^r G = 0, \label{eq:star}
\end{equation} 
then (\ref{x_estmate_no_z}) provides an immediate reconstruction of the state of (\ref{plant}) for all $t\geq0$.

The proof follows the calculations (\ref{observer1}) -- (\ref{x_estmate}).

The convergence speed of $\hat{x}(t)$ to $x(t)$ depends on the eigenvalues of the matrix $F$. Therefore, we can choose the matrix $F$ using pole placement procedure to satisfy desired transients. 

\textbf{Remark 2.} Direct computations show that in the case $F^r G + L = 0$ we get $\tilde{x}(0)=-z_r(0)$.

\section{Estimation of unknown parameters and state vector}

The observer proposed in the previous section provides a finite-time estimation of the state vector benefiting from the derivatives of the output. In the current section we will demonstrate how this observer can be used to estimate unknown parameters and disturbances, as well as to further estimate the state vector without using derivatives of the output signal.

\subsection{Estimation of unknown parameters}
Let us show how the vector of unknown parameters can be evaluated together with the harmonic disturbance. For brevity we perform the main computations for the case $\varphi(t, x, y)= x(t)$ only. The other cases can be resolved in a completely analogous manner. Assuming that the condition (\ref{eq:star}) is verified and by substituting the expression (10) for the estimation of $x(t)$ into (\ref{plant}) and expression of solution for $\theta(t)$, we obtain
\begin{equation}
\label{aux1}
    \begin{split}
        &\dot{z}_r + F^{r-1}G\dot{y}+\ldots+Gy^{(r)}=A(z_r+F^{r-1}Gy+\ldots+Gy^{(r-1)})+\\
        &+B(z_r+F^{r-1}Gy+\ldots+Gy^{(r-1)})^THe^{\Gamma t}\xi(0)+Bf(t),
    \end{split}    
\end{equation}
 
Equation (\ref{aux1}) involves unmeasured derivatives of the output signal, an unknown external disturbance $f(t)$, and the initial conditions $\xi(0)$. Let us apply $r$-th order linear filters and the swapping lemma \cite{pyrkin2019adaptive} which will allow us to eliminate the terms containing unmeasured derivatives in the regressor for $\xi(0)$. To illustrate this approach, and simplifying the writing, consider the case when the relative degree is equal two and the disturbance is represented by a  harmonic signal. Equation (\ref{aux1}) takes the form
\begin{equation}
\label{aux2}
    \begin{split}
        &\dot{z}_2 +FG\dot{y}+G\ddot{y}=A(z_2+FGy+G\dot{y})+\\    &+B(z_2+FGy+G\dot{y})^THe^{\Gamma t}\xi(0)+Bf.
    \end{split}
\end{equation}

Apply to (\ref{aux2}) linear filter $\frac{\lambda^2_2}{(p+\lambda_2)^2}$ and swapping lemma
\begin{equation}
\label{swap}
    \begin{split}
        & \frac{\lambda_2}{p+\lambda_2}\left[\dot{y}G^THe^{\Gamma t}\xi(0) \right]= G^THe^{\Gamma t}\xi(0)\frac{\lambda_2p}{p+\lambda_2}[y]-\\
        &  - \frac{1}{p+\lambda_2}\left[G^TH\Gamma e^{\Gamma t}\xi(0)\frac{\lambda_2p}{p+\lambda_2}[y] \right]   =\left[\frac{\lambda_2p}{p+\lambda_2}[y]G^THe^{\Gamma t}-\right.\\
        & \left.- \frac{1}{p+\lambda_2}\left[\frac{\lambda_2p}{p+\lambda_2}[y]G^TH\Gamma e^{\Gamma t} \right]  \right] \xi(0) =S_1(t)\xi(0),
    \end{split}
\end{equation}
where
$$
S_1(t)= \frac{\lambda_2p}{p+\lambda_2}[y]G^THe^{\Gamma t}-\frac{1}{p+\lambda_2}\left[\frac{\lambda_2p}{p+\lambda_2}[y]G^TH\Gamma e^{\Gamma t} \right].
$$
Rewrite the equation (\ref{aux2}) as follows
$$
    q_2(t)=B\left(S_0(t)+\frac{\lambda_2}{p+\lambda_2}[S_1(t)] \right)\xi(0)+B\bar{f}(t),
$$
where
\[q_2(t)=\frac{\lambda_2^2}{(p+\lambda_2)^2}\left[(\dot{z}_2+FG\dot{y}+G\ddot{y})-A(z_2+FGy+G\dot{y}) \right], \]
\[S_0(t)=\frac{\lambda_2^2}{(p+\lambda_2)^2}[(z_2 + FGy)^THe^{\Gamma t}],\; \bar{f}(t)=\frac{\lambda_2^2}{(p+\lambda_2)^2}[f(t)].\]
Thus, by applying a filter to (\ref{aux1}), and after a series of transformations, we obtain the following expression for a system with an arbitrary relative degree:
\begin{equation}
\label{filter-and-swapping}
    \begin{split}
        & q_r(t)= B\left[S_0(t)+\frac{\lambda^{r-1}}{(p+\lambda)^{r-1}}[S_1(t)]+\ldots + S_{r-1}(t) \right]\xi(0)+ B\bar{f}(t),
    \end{split}
\end{equation}
where 
$$
\begin{aligned}
  &q_r(t)=\frac{\lambda^r}{(p+\lambda)^r}\left[ \left( \dot{\bar{z}}_r + F^{r-1}G\dot{y} + \ldots + Gy^{(r)} \right) \right.- \\
  &\left. A \left( \bar{z}_r + F^{r-1}Gy + \ldots + Gy^{(r-1)} \right]) \right]=\\
  & =\frac{\lambda^rp}{(p+\lambda)^r}\bar{z}_r + F^{r-1}G\frac{\lambda^rp}{(p+\lambda)^r}y + \hdots + G\frac{(\lambda p)^r}{(p+\lambda)^r}y-\\
  & - A\left(\frac{\lambda^r}{(p+\lambda)^r} \bar{z}_r + F^{r-1}G \frac{\lambda^r}{(p+\lambda)^r} y + \hdots + G \frac{\lambda^rp^{r-1}}{(p+\lambda)^r}y \right)
\end{aligned}
$$
$$S_0(t)=\frac{\lambda^r}{(p+\lambda)^r} \left[ (\bar{z}_r + F^{r-1}Gy)^THe^{\Gamma t} \right],\; \bar{f}(t)=\frac{\lambda^r}{(p+\lambda)^r}[f(t)].$$
The resulting equation in a simplified form is as follows:
\begin{equation}
    \label{aux_simplified}
    q_r(t)=B\bar{S}_r(t)\xi(0)+B\bar{f}(t),
\end{equation}
where 
$$
\bar{S_r}(t)=S_0(t)+\frac{\lambda^{r-1}}{(p+\lambda)^{r-1}}[S_1(t)]+\ldots + S_{r-1}(t).
$$

Next, we apply the properties of a sinusoidal signal:
$$
p^2\bar{f}(t)=-\omega^2\bar{f}(t),
$$
where $\omega$ is an unknown frequency. Rewrite equation (\ref{aux_simplified})

\begin{equation}
\label{filtered_aux_simple}
    p^2[q_r(t)-B\bar{S}_r(t)\xi(0)] = -\omega^2[q_r(t)-B\bar{S}_r(t)\xi(0)].
\end{equation}
Multiply (\ref{filtered_aux_simple}) by row-vector $\bar{B}$ such that $\bar{B}B=1$. Then, apply a second order filter $\frac{\lambda_r^2}{(p+\lambda_r)^2}$ ($\lambda_r > 0$) and group the unknown terms to obtain a regression
\begin{equation}
\label{autoregression}
    q_r^*(t) = m_r^T(t)k_r,
\end{equation}
where  $q_r^*(t)=\frac{\lambda_r^2p^2}{(p+\lambda_r)^2}[\bar{B}q_r(t)]$,
$$m_r^T(t)=\left[ \frac{\lambda_r^2p^2}{(p+\lambda_r)^2}[\bar{S}_r(t)], \frac{\lambda_r^2}{(p+\lambda_r)^2}[-\bar{B}q_r(t)],\frac{\lambda_r^2}{(p+\lambda_r)^2}[\bar{S}_r(t)] \right],$$
$$k_r=[\xi(0);\omega^2;\omega^2\xi(0)].$$
If the signal $\bar{f}(t)$ includes several harmonics, then the system (\ref{aux_simplified}) can also be transformed to a linear regression \cite{multiharm2reg}.

Equation (\ref{autoregression}) has the form of linear regression and enables the estimation of the initial conditions for the vector $\hat{\xi}(0)$ and disturbance frequency $\hat{\omega}$. Transients can exhibit both asymptotic (using the gradient descent method) and finite-time convergence. We propose to use the DREM method based on Kreisselmeier’s regressor extension scheme described in \cite{pyrkin2019adaptive} (which does not destroy the excitation in the system \cite{Aranovskiy2023}). Let us introduce the vector $\Phi_r(t)$ and the matrix $Y_r(t)$ as solutions of the differential equations
\begin{align}
    \label{kreiselmer_2}
    \begin{split}
\dot \Phi _r(t) =  - {h_r}{\Phi _r}(t) + {m_r}(t)m_r^T(t), \Phi _r(t_0)=0,\\
\dot Y_r(t) =  - {h_r}{Y_r}(t) + {m_r}(t){q_r}(t), Y_r(t_0)=0,\\
    \end{split}
\end{align}
where $h_r>0$ is a tuning coefficient.
Let us denote the adjoint matrix of $\Phi_r(t)$ as $\mathrm{Adj}(\Phi{_r}(t))$. After applying DREM, we obtain
$$
    \Upsilon_{r}(t)=\Delta_r(t)k_{r},
$$
where $\Delta_r(t)=\mathrm{det}(\Phi_r(t)), \Upsilon_{r}(t)=\mathrm{Adj}(\Phi_r(t))Y_r(t)=[\Upsilon_{r1}(t)\dots\Upsilon_{r, \dim{k_r}}(t)]^\top$. Therefore, we can determine each element of the vector $k_r$ using the following equations \cite{Korotina2024}:
\begin{equation}
\label{dir_DREM_2}
\hat k_{ri}^{dir}(t) = \frac{{{\Upsilon _{ri}}(t)}}{{{\rm{ max\{ }}{\Delta _r}(t),\varepsilon_r \} }},i = \overline {1,\dim{k_r}},  
\end{equation}
where $\varepsilon_r>0$ is a small constant incorporated to guarantee the practicality of executing the direct estimation for the initial transients.

After applying conventional noise filters to equation (\ref{dir_DREM_2}) allows us to benefit the new measurements to improve the accuracy of estimation \cite{Korotina2024} in the presence of disturbances. We have the following result.

\textbf{Proposition 1.} If $\Delta_r\notin L_2$, then the algorithm (\ref{autoregression})-(\ref{dir_DREM_2}) provides an estimation of the initial conditions $\xi(0)$ and the frequency of the external disturbance $\omega$ in a finite time.

Proof follows the above calculations (\ref{swap}) -- (\ref{dir_DREM_2}).

\textbf{Remark 3.} In the case of $\varphi(x,y,t)$ is a nonlinear function of the form \[\varphi (x,y,t) = \alpha (y)x(t),\]
where $\alpha(y)$ is a known function, satisfying condition  $\dot \alpha (y) = \frac{{\partial \alpha }}{{\partial y}}\dot y = \beta \dot{y}$ with a constant $\beta$, instead of (21) we obtain
\[\begin{array}{l}
\dot{\bar{z}}_r + F^{r-1}G\dot{y}+\ldots+Gy^{(r)}=\\
A(\bar{z}_r+F^{r-1}Gy+\ldots+Gy^{(r-1)})+ Bf(t)\\
+B\alpha(y)(\bar{z}_r+F^{r-1}Gy+\ldots+Gy^{(r-1)})^THe^{\Gamma t}\xi(0).
\end{array}\]
Application of swapping lemma allows to solve problem entirely analogous to the case $\varphi=x(t)$ as follows:
\[\begin{array}{l}
\frac{\lambda }{{p + \lambda }}[\alpha (y)\dot y{G^T}H{e^{\Gamma t}}\xi (0)] = \alpha (y){G^T}H{e^{\Gamma t}}\xi (0)\frac{{p\lambda }}{{p + \lambda }}[y] - \\
\frac{1}{{p + \lambda }}\left[ {(\beta y{G^T}H{e^{\Gamma t}} + \alpha (y){G^T}H\Gamma {e^{\Gamma t}})\xi (0)\frac{{p\lambda }}{{p + \lambda }}[y]} \right].
\end{array}\]

From equation (\ref{aux_simplified}), taking into account Theorem 2, we obtain the following
\begin{equation}
\label{f-regression}
    \begin{split}
         & \bar{f}(t) = \bar{B}q_r(t) - \bar{S}_r(t)\xi(0) = a_1 \sin (\omega t) + a_2 \cos (\omega t) = \psi^T(t) a,
    \end{split}   
\end{equation}
where $\psi^T(t)=[\sin (\omega t), \cos (\omega t)]$, $a = [a_1; a_2].$

Then the linear regression equation (\ref{f-regression}) can be solved using the DREM method based on Kreisselmeier’s regressor extension scheme to find the amplitude of the disturbance. Similar to the adaptive algorithm (\ref{autoregression})-(\ref{dir_DREM_2})  the amplitude is determined as follows

\begin{equation}
\label{a-regression1}
   \Upsilon_{a}(t)=\Delta_a(t)a,
\end{equation}

\begin{equation}
\label{a-regression2}
\hat a_{i}^{dir}(t) = \frac{{{\Upsilon _{ai}}(t)}}{{{\rm{ max\{ }}{\Delta _a}(t),\varepsilon_a \} }}, i=\overline{1, \dim a},
\end{equation}
where $\Delta_a(t)=\mathrm{det}(\Phi_a(t)), \Upsilon_{a}(t)=\mathrm{Adj}(\Phi_a(t))Y_a(t)=[\Upsilon_{a1}(t)\dots\Upsilon_{a, \dim{a}}(t)]^\top$, $\Phi_a(t)$ and $Y_a(t)$ are solutions of the differential equations
$$
    \begin{aligned}
\dot \Phi _a(t) =  - {h_a}{\Phi _a}(t) + {\psi}(t)\psi^T(t), \Phi _a(t_0)=0,\\
\dot Y_a(t) =  - {h_a}{Y_a}(t) + {\psi}(t){\bar{f}}(t), Y_a(t_0)=0,\\
    \end{aligned}
$$
where $\varepsilon_a>0$ and $h_a>0$ are tuning parameters

On the other hand, considering the determined value of the amplitude in (\ref{a-regression1})-(\ref{a-regression2}) together with (\ref{filter-and-swapping}), it is evident that the external disturbance $f(t)$ can be reconstructed as follows
\begin{equation}\label{36}
    f(t)=\frac{(p+\lambda)^r}{\lambda^r}[\bar{f}(t)]
\end{equation}
since with known parameters of $\bar{f}(t)$ its derivatives can be calculated due to the harmonic nature of the signal. 

\textbf{Proposition 2.} If $\Delta_a\notin L_2$, then the algorithm (\ref{f-regression}) -- (\ref{36}) provides an estimation of the disturbance $f(t)$ in a finite time.


 \textbf{Remark 4.} Proposed algorithm allows the parameter vector to be estimated even in the case when the regressor depends on the $x(t)$. It is the main difference between the solution proposed in this work and approach \cite{2pyrkin2023adaptive}  that is applicable if the function $\varphi$ depends only on the output signal $y(t)$. 

\subsection{State estimation}
Estimates of all unknown parameters allow us to construct an arbitrary observer of the state vector for (\ref{plant}). Introduce the following state vector observer that does not require derivatives of the output signal
\begin{equation}
    \label{obs_fin}
        \begin{aligned}
            &\dot{\bar{x}} = A\bar{x}+B[u+\alpha(y,t)\bar{x}^T\hat{\theta}
            +\hat{f}]+Ky-KC\bar{x},\\
        \end{aligned}
\end{equation}
where $K$ is a design matrix. Taking into account 
$x^T\theta=(\bar{x}+\tilde{x})^T(\hat{\theta}+\tilde{\theta})$,  obtain observation error dynamic model
\begin{equation}
    \label{err_dyn_fin1}
    \begin{aligned}
        &\dot{\tilde{x}} = (A-KC)\tilde{x} + B\tilde{f} + \alpha(y,t)Bx^T\tilde{\theta} + \alpha(y,t)B\tilde{x}^T\hat{\theta}=\\
        & = (A-KC)\tilde{x} + \alpha(y,t)B\hat{\theta}^T\tilde{x} +B\delta,
    \end{aligned}
\end{equation}
where $\delta = \tilde{f} +\alpha(y,t)x^T\tilde{\theta}$ vanishes in a finite time.
Choose $K=NC^T$, where $N=N^T>0$ is a solution of Riccati equation
\begin{equation}
\label{ricatti}
\dot{N}(t) = 2\gamma (t) N(t) +N(t)A^T +AN(t) - 2N(t)C^TCN(t) + \mu (t)BB^T,
\end{equation}
where $\gamma$ and $\mu$ are positive functions of time defined below.\\
Introduce Lyapunov function $V=\tilde{x}^TP\tilde{x}$, where $P=N^{-1}$, and consider its derivative
$$
\begin{aligned}
& \dot{V}=(\delta B^T - \tilde{x}^TC^TK^T +\alpha(y,t)\tilde{x}^T\hat{\theta}B^T + \tilde{x}^TA^T)P\tilde{x} + \tilde{x}^T\dot{P}\tilde{x} + \\
& + \tilde{x}^TP(A\tilde{x} + \alpha(y,t)B\hat{\theta}^T\tilde{x}-KC\tilde{x}+B\delta).
\end{aligned}
$$
Since $PN=I$, $\dot{P}N + P\dot{N} = 0$ we obtain $\dot{P} = -P\dot{N}P$, then
$$
\begin{aligned}
&\dot{V} = -\tilde{x}^TP\dot{N}P\tilde{x} + \tilde{x}^T(A^TP+PA-2CC^T)\tilde{x} +2\delta B^TP\tilde{x}+2\alpha(y,t)\tilde{x}^TPB\hat{\theta}^T\tilde{x}=\\
&=-\tilde{x}^TP(2\gamma N + NA^T + AN-2NC^TCN+\mu BB^T)P\tilde{x} +2\delta B^TP\tilde{x}+\\
& + \tilde{x}^TP(NA^T+AN-2NC^TCN)P\tilde{x} +2\alpha(y,t)\tilde{x}^TPB\hat{\theta}^T\tilde{x} = -2\gamma\tilde{x}^TP\tilde{x}-\\
& - (\tilde{x}^TC^T)^2 -\mu \tilde{x}^TPBB^TP\tilde{x}+2\delta B^TP\tilde{x} +2\alpha(y,t)\tilde{x}^TPB\hat{\theta}^T\tilde{x}  \\
& \le -2\gamma\tilde{x}^TP\tilde{x} - \mu(\tilde{x}^TPB)^2 + \delta^2 +(\tilde{x}^TPB)^2 + k(\hat{\theta}^TN\hat{\theta})(\tilde{x}^TPB)^2+\frac{1}{k}\alpha^2(y,t)\tilde{x}^TP\tilde{x},
\end{aligned}
$$
where $k>0$ and the Young's inequality was used on the last step. \\
Choose $\mu \ge 1+k(\hat{\theta}^TN\hat{\theta})$ and $\gamma \ge \frac{\alpha^2(y,t)}{k}$. Then the derivative of the Lyapunov function is bounded by the inequality
$$
\dot{V} \le -\gamma V + \delta^2.
$$
Since $\delta = 0$ in a finite time, the observer (\ref{obs_fin}) provides convergence of the $\tilde{x}$ to zero. Consequently, we can formulate our main result, whose proof is given above:

\textbf{Theorem 3.} Under Assumptions 1--5, the observers of propositions 1 and 2 guarantee the estimation of $\theta(t)$ and $f(t)$ in a finite time, while the time-varying observer \eqref{obs_fin} asymptotically reconstructs the state.


\section{Simulation}
To demonstrate the functionality and effectiveness of the proposed approach, let us consider a second-order dynamic system with a relative degree $r=2$, given by
\begin{align}
\label{example}
    \left\{ \begin{array}{*{35}{l}}
    \dot{x}_1(t) = x_2(t)+x_1(t)\theta_1(t),\\
     \dot{x}_2(t) =-x_1(t)-2x_2(t)+u(t)+x_2(t)\theta_2(t)+f(t),\\
    y(t)=x_1(t)+\varsigma(t),\\
\end{array}\right.
\end{align}
with $f(t)=5\sin(2t)$, $x(0)=\left[ {\begin{array}{*{20}{c}}
-2\\
2
\end{array}} \right]$, parameter $\theta(t)$ is generated by the output of the linear generator with matrices $H = \left[ {\begin{array}{*{20}{c}}
2&0\\
3&0
\end{array}} \right]$, $\Gamma  = \left[ {\begin{array}{*{20}{c}}
0&1\\
{ - 36}&0
\end{array}} \right]$ with initial conditions  $\xi (0) = \left[ {\begin{array}{*{20}{c}}
{ - 1}\\
{ - 2}
\end{array}} \right]$ and $\varsigma(t)$ is measurement noise with normal distribution, mean 0.01, variance 0.001, which is introduced to demonstrate robustness of the proposed approach.\\

Let us rewrite the equation (\ref{example}) in the form \eqref{plant}:
\[\left\{ \begin{array}{l}
\dot x(t) = \left[ {\begin{array}{*{20}{c}}
0&1\\
{ - 1}&{ - 2}
\end{array}} \right]x(t) + \left[ {\begin{array}{*{20}{c}}
0\\
1
\end{array}} \right][u(t)+ \left[ {\begin{array}{*{20}{c}}
{{\theta _1}(t)}&{{\theta _2}(t)}
\end{array}} \right]x(t) + f(t)]\\
y(t) = \left[ {\begin{array}{*{20}{c}}
1&0
\end{array}} \right]x(t)
\end{array} \right.\]
then the state observer (\ref{x_estmate}) can be applied. Matrix $F$ is chosen to be Hurwitz,
$$G = B{[{(CAB)^T}CAB]^{ - 1}}{(CAB)^T} 
= \left[ {\begin{array}{*{20}{c}}
0\\
1
\end{array}} \right],$$ 
$$M = A - GC{A^2} = \left[ {\begin{array}{*{20}{c}}
0&1\\
0&0
\end{array}} \right],$$ 
$$L = place({M^T},{C^T},{[\begin{array}{*{20}{c}}
{ - 15}&{ - 10]}
\end{array})^T},$$
$${F = M - LC} = \left[ {\begin{array}{*{20}{c}}
-25&1\\
{ - 125}&0
\end{array}} \right],$$ 
with initial conditions of the observer ${z_2}(0) = \left[ {\begin{array}{*{20}{c}}
{ - 0.5}\\
0.5
\end{array}} \right]$. The matrix $\bar{B}$ is chosen as 
$\left[ {\begin{array}{*{20}{c}}
1&1 \end{array}} \right]$.

\begin{figure}
    \centering
    \includegraphics[width=0.6\linewidth]{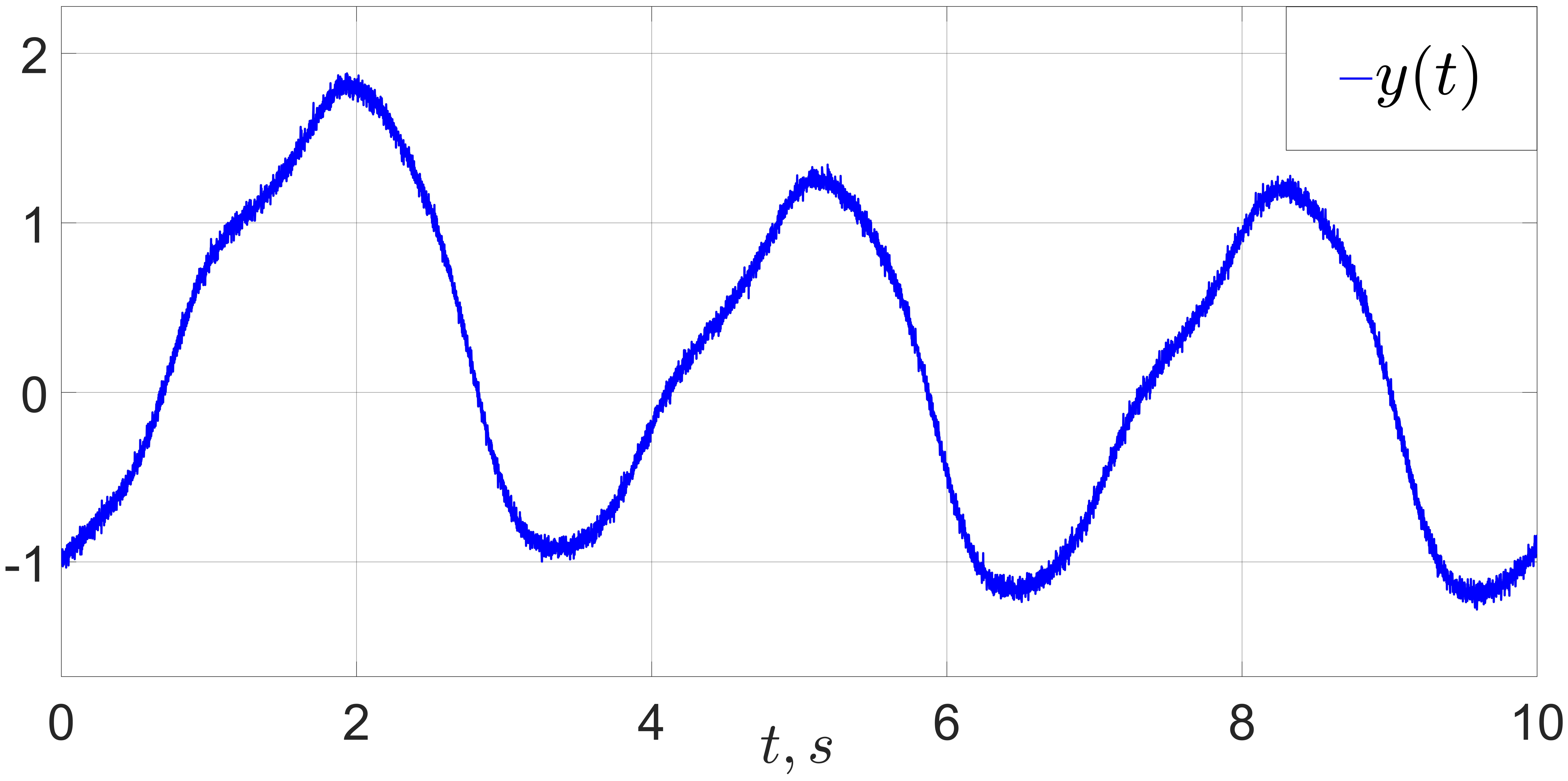}
    \caption{The output signal with measurement noise}
    \label{fig_2_output}
\end{figure}

We choose the filters as follows
\[\frac{{{\lambda}}}{{p + {\lambda}}}=\frac{{{\lambda _1}}}{{p + {\lambda _1}}} = \frac{{{\lambda _2}}}{{p + {\lambda _2}}} = \frac{{{\lambda _r}}}{{p + {\lambda _r}}} =  \ldots  = \frac{5}{{p + 5}}.\]For the Kreisselmeier’s regressor extension scheme, the tuning parameter  are selected as\\
\[h_r=\ldots=h_a=0.5.\]
The direct estimation can be sensitive to noise at the initial time due to the small value of $\Delta_r, \Delta_a$. Therefore, the parameters $\varepsilon_r=\varepsilon_a$ is chosen to be $10^{-3}$. In the simulation a low-pass filter is applied to \eqref{dir_DREM_2} and \eqref{a-regression2} with a parameter $\sigma=0.7$.

The output signal $y(t)$ measured under the noise $\varsigma(t)$ is  shown on the Figure \ref{fig_2_output}. Denote $\hat{\tilde {x}}(t)$ and  $\tilde{f}(t)=f(t)-\hat{f}(t)$ the errors of the initial observer error and the external disturbance estimation error, respectively.

The transient processes  of simulation of algorithms  (\ref{autoregression})-(\ref{dir_DREM_2}) and (\ref{f-regression})-(\ref{a-regression2}) are shown in the Figures 2, 3.

\begin{figure}
    \centering
    \includegraphics[width=0.6\linewidth]{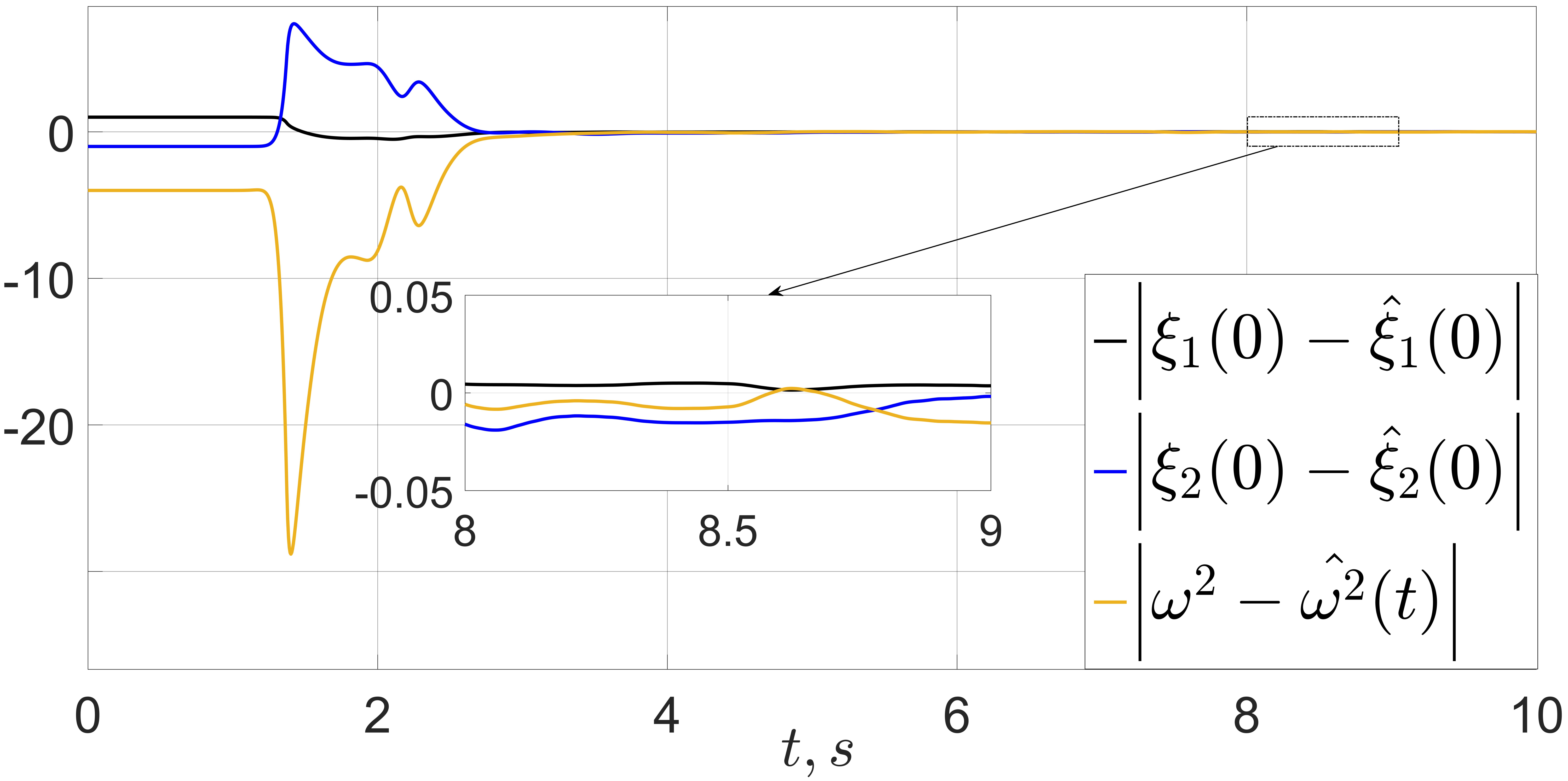}
    \caption{The error in estimating the initial conditions of the time-varying parameter $\xi(0)$ and the external disturbance frequency $\omega(t)$}
    \label{fig_chung_2}
\end{figure}
\begin{figure}
    \centering
    \includegraphics[width=0.6\linewidth]{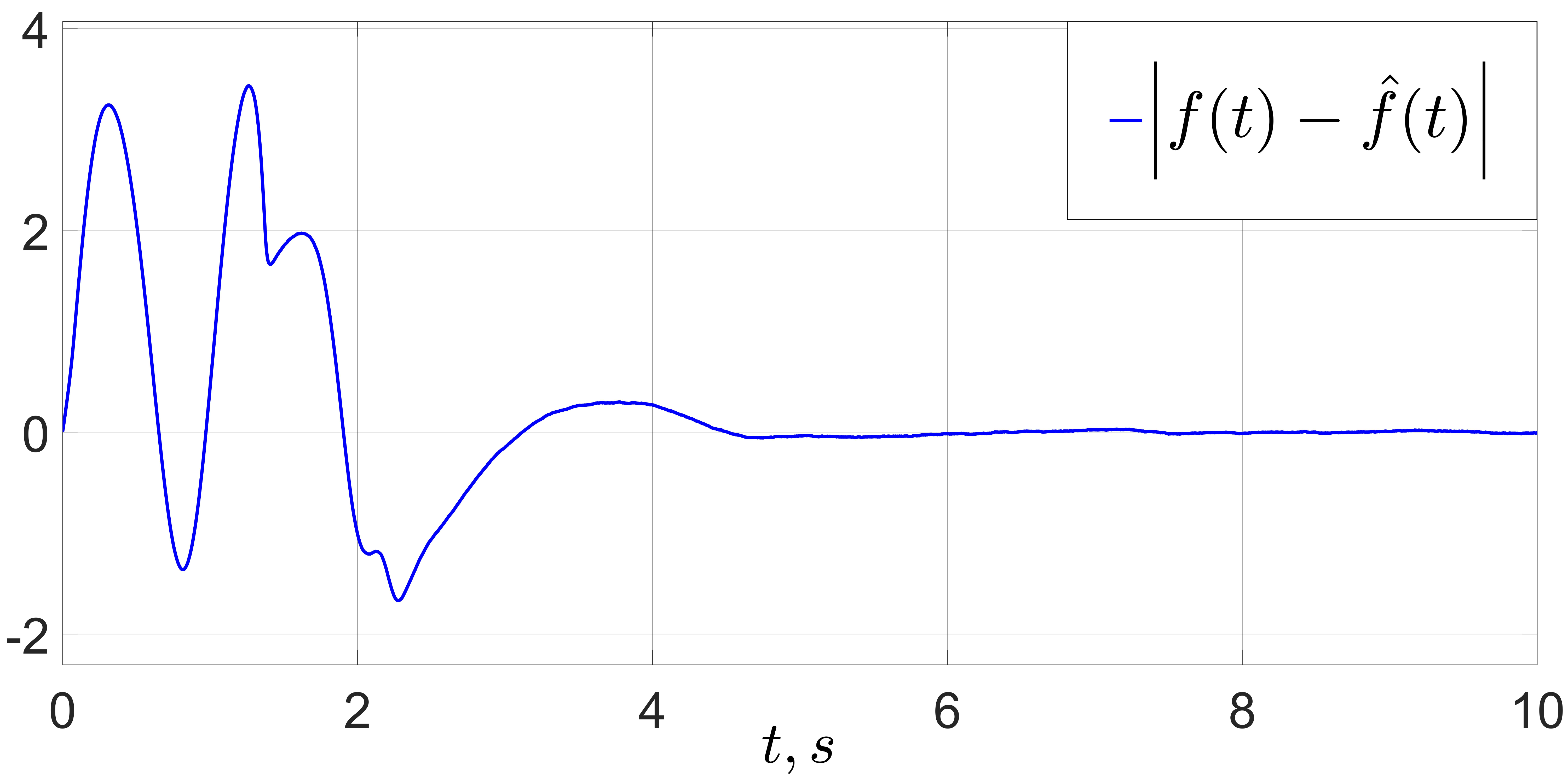}
    \caption{The error in estimating the external disturbance $f(t)$}
    \label{fig_disturbawnce_2}
\end{figure}
\textbf{Remark 6.}  To reduce the computational complexity it is possible to fix the value of  $\hat{\tilde x}(0)$ using the modified DREM algorithm \cite{38ortega2020new} as a constant for the predetermined time $t_D$. Therefore, it is possible to provide convergence with a predetermined time.

\begin{figure}
    \centering
    \includegraphics[width=0.7\linewidth]{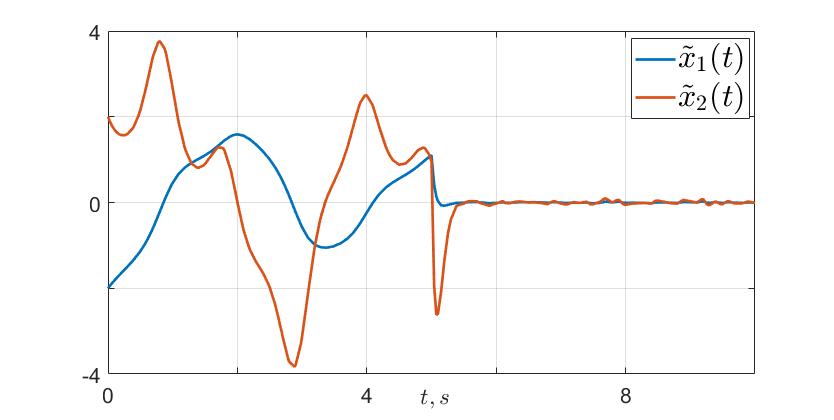}
    \caption{The state observation error $\tilde{x}(t)$}
    \label{fig_obs_error}
\end{figure}

To construct a state observer, it is necessary to choose the matrix $K$ to ensure the stability of the closed-loop system (25). In the general case, this requires solving the Ricatti equation (\ref{ricatti}). In the simple example under consideration with bounded $\theta$ and $\alpha(y)$, we can choose
$$
K=\left[
\begin{split}
    &23\\
    &103
\end{split}
\right].
$$
Figure \ref{fig_obs_error} shows the transient processes of the state vector observation error. The observer is started after finite-time estimation of the unknown parameters and disturbance. Minor deviations of the error from zero are caused by the presence of measurement noise.

\section{Conclusion}
The paper addressed the problem of unknown input observer synthesis for a class of time-varying uncertain systems under external multiharmonic disturbances. The proposed method provides state vector estimation for systems with arbitrary relative degrees. Additionally, obtained solution allows identification of time-varying parameters and disturbances. All estimates can be derived in a finite time. In the future, the utilization of the proposed unknown input observer with arbitrary relative degrees can be extended to more complex problems, such as systems with unknown nonlinear time varying parameters or nonlinear parameter dependence on the state vector.



 \bibliographystyle{elsarticle-num} 
 \bibliography{uio_ver1}

\begin{thebibliography}{10}
\expandafter\ifx\csname url\endcsname\relax
  \def\url#1{\texttt{#1}}\fi
\expandafter\ifx\csname urlprefix\endcsname\relax\def\urlprefix{URL }\fi
\expandafter\ifx\csname href\endcsname\relax
  \def\href#1#2{#2} \def\path#1{#1}\fi

\bibitem{kailath1980linear}
T.~Kailath, Linear systems, Vol. 156, Prentice-Hall Englewood Cliffs, NJ, 1980.

\bibitem{2}
J.~O'Reilly, Observers for linear systems, Vol. 170, Academic press, 1983.

\bibitem{3}
B.~Liu, J.~Si, Fault isolation filter design for linear time-invariant systems, IEEE Transactions on Automatic Control 42~(5) (1997) 704--707.

\bibitem{4}
D.~G. Luenberger, Observing the state of a linear system, IEEE transactions on military electronics 8~(2) (1964) 74--80.

\bibitem{linearization5}
A.~Banaszuk, Approximate feedback linearization of nonlinear control systems, Georgia Institute of Technology, 1995.

\bibitem{linearization7}
W.~Li, E.~Todorov, Iterative linearization methods for approximately optimal control and estimation of non-linear stochastic system, International Journal of Control 80~(9) (2007) 1439--1453.

\bibitem{linearization9}
L.~Socha, T.~Soong, Linearization in analysis of nonlinear stochastic systems, 1991.

\bibitem{approximation6}
E.~Y. Butyrsky, Piecewise linear approximation in the problem of filtering and signal detection, National Security and Strategic Planning~(1) (2021) 34--43.

\bibitem{approximate8}
J.~Deutscher, C.~Schmid, A state space embedding approach to approximate feedback linearization of single input nonlinear control systems, International Journal of Robust and Nonlinear Control 16~(9) (2006) 421--440.

\bibitem{Kalman13}
A.~Barrau, S.~Bonnabel, {The invariant extended Kalman filter as a stable observer}, IEEE Transactions on Automatic Control 62~(4) (2016) 1797--1812.

\bibitem{kalman14}
Y.~Song, J.~W. Grizzle, {The extended Kalman filter as a local asymptotic observer for nonlinear discrete-time systems}, in: 1992 American control conference, IEEE, 1992, pp. 3365--3369.

\bibitem{kalman15}
A.~Valibeygi, K.~Vijayaraghavan, et~al., {A comparative study of extended Kalman filter and an optimal nonlinear observer for state estimation}, in: 2017 American Control Conference (ACC), IEEE, 2017, pp. 5211--5216.

\bibitem{sliding16}
A.~Zuev, A.~Zhirabok, V.~Filaretov, A.~Protsenko, Identification of defects in non-stationary systems based on moving observers, Mechatronics, automation, control 22~(12) (2021) 625--633.

\bibitem{sliding17}
B.~Andrievsky, A.~L. Fradkov, et~al., Adaptive aircraft control with sliding mode identification, Management of large systems: collection of works~(26) (2009) 113--144.

\bibitem{5_bobtsov2022adaptive}
A.~Bobtsov, N.~Nikolaev, O.~Slita, O.~Kozachek, O.~Oskina, {Adaptive observer for a LTV system with partially unknown state matrix and delayed measurements}, in: 2022 14th International Congress on Ultra Modern Telecommunications and Control Systems and Workshops (ICUMT), IEEE, 2022, pp. 165--170.

\bibitem{6_bobtsov2021state}
A.~Bobtsov, N.~Nikolaev, R.~Ortega, D.~Efimov, {State observation of LTV systems with delayed measurements: A parameter estimation-based approach with fixed convergence time}, Automatica 131 (2021) 109674.

\bibitem{35lyung1991identification}
L.~Ljung, System Identification Theory for the User, Science, 1991.

\bibitem{36vediakova2020drem}
A.~Vediakova, A.~Vedyakov, A.~Bobtsov, A.~Pyrkin, {DREM-based parametric estimation of bias-affected damped sinusoidal signals}, in: 2020 European Control Conference (ECC), IEEE, 2020, pp. 214--219.

\bibitem{37ovcharov2020finite}
A.~Ovcharov, A.~Pyrkin, A.~Bobtsov, D.~Bazylev, R.~Ortega, A.~Vedyakov, {Finite time observer for induction motors based on DREM algorithm}, in: 2020 European Control Conference (ECC), IEEE, 2020, pp. 1318--1323.

\bibitem{38ortega2020new}
R.~Ortega, S.~Aranovskiy, A.~A. Pyrkin, A.~Astolfi, A.~A. Bobtsov, {New results on parameter estimation via dynamic regressor extension and mixing: Continuous and discrete-time cases}, IEEE Transactions on Automatic Control 66~(5) (2020) 2265--2272.

\bibitem{2pyrkin2023adaptive}
A.~Pyrkin, A.~Bobtsov, R.~Ortega, A.~Isidori, An adaptive observer for uncertain linear time-varying systems with unknown additive perturbations, Automatica 147 (2023) 110677.

\bibitem{40hou1992design}
M.~Hou, P.~C. Muller, Design of observers for linear systems with unknown inputs, IEEE Transactions on automatic control 37~(6) (1992) 871--875.

\bibitem{41chen1996design}
J.~Chen, R.~J. Patton, H.-Y. Zhang, Design of unknown input observers and robust fault detection filters, International Journal of control 63~(1) (1996) 85--105.

\bibitem{42warrad2018design}
S.~B. Warrad, O.~Boubaker, Design of unknown input observers for linear systems with state and input delays, in: 2018 15th International Multi-Conference on Systems, Signals \& Devices (SSD), IEEE, 2018, pp. 1--5.

\bibitem{1982observer}
N.~Kobayashi, T.~Nakamizo, An observer design for linear systems with unknown inputs, International Journal of Control 35~(4) (1982) 605--619.

\bibitem{50warrad2018design}
S.~B. Warrad, O.~Boubaker, Design of unknown input observers for linear systems with state and input delays, in: 2018 15th International Multi-Conference on Systems, Signals \& Devices (SSD), IEEE, 2018, pp. 1--5.

\bibitem{51sharma2016unknown}
V.~Sharma, V.~Agrawal, B.~Sharma, R.~Nath, Unknown input nonlinear observer design for continuous and discrete time systems with input recovery scheme, Nonlinear Dynamics 85 (2016) 645--658.

\bibitem{57coutinho2022sufficient}
P.~H. Coutinho, I.~Bessa, W.-B. Xie, A.-T. Nguyen, R.~M. Palhares, A sufficient condition to design unknown input observers for nonlinear systems with arbitrary relative degree, International Journal of Robust and Nonlinear Control 32~(15) (2022) 8331--8348.

\bibitem{60aguilera2012sensor}
A.~Aguilera-Gonzalez, D.~Theilliol, M.~Adam-Medina, C.~Astorga-Zaragoza, M.~Rodrigues, {Sensor Fault and Unknown Input Estimation Based on Proportional Integral Observer Applied to LPV Descriptor Systems.}, IFAC Proceedings Volumes 45~(20) (2012) 1059--1064.

\bibitem{56gomez2019sensor}
S.~G{\'o}mez-Pe{\~n}ate, G.~Valencia-Palomo, F.-R. L{\'o}pez-Estrada, C.-M. Astorga-Zaragoza, R.~A. Osornio-Rios, I.~Santos-Ruiz, {Sensor fault diagnosis based on a sliding mode and unknown input observer for Takagi-Sugeno systems with uncertain premise variables}, Asian Journal of Control 21~(1) (2019) 339--353.

\bibitem{62ichalal2015unknown}
D.~Ichalal, S.~Mammar, {On unknown input observers for LPV systems}, IEEE Transactions on Industrial Electronics 62~(9) (2015) 5870--5880.

\bibitem{63floquet2007sliding}
T.~Floquet, C.~Edwards, S.~K. Spurgeon, On sliding mode observers for systems with unknown inputs, International Journal of Adaptive Control and Signal Processing 21~(8-9) (2007) 638--656.

\bibitem{58wang2021novel}
X.~Wang, C.~P. Tan, L.~Liu, Q.~Qi, A novel unknown input interval observer for systems not satisfying relative degree condition, International Journal of Robust and Nonlinear Control 31~(7) (2021) 2762--2782.

\bibitem{farza2009adaptive}
M.~Farza, M.~M’Saad, T.~Maatoug, M.~Kamoun, {Adaptive observers for nonlinearly parameterized class of nonlinear systems}, Automatica 45~(10) (2009) 2292--2299.

\bibitem{bicycle_model}
J.~Ackermann, T.~Bunte, W.~Sienel, H.~Jeebe, K.~Naab, {Driving safety by robust steering control}, Proc. Int. Symp. Adv. Veh. Control (1996) 377--394.

\bibitem{ship_model}
R.~Skjetne, O.~Smogeli, T.~I. Fossen, {Modeling, identification and adaptive maneuvering of Cybership II: A complete design with experiments}, Proc. IFAC Conf. Control Appl. Mar. Syst. (2004) 203--208.

\bibitem{pyrkin2019adaptive}
A.~Pyrkin, A.~Bobtsov, R.~Ortega, A.~Vedyakov, S.~Aranovskiy, Adaptive state observers using dynamic regressor extension and mixing, Systems \& Control Letters 133 (2019) 104519.

\bibitem{multiharm2reg}
T.~N. Khac, S.~Vlasov, A.~Pyrkin, {Parameters estimation of multi-sinusoidal signal in finite-time}, Cybernetics and Physics 11~(2) (2022) 74--81.

\bibitem{Aranovskiy2023}
S.~Aranovskiy, R.~Ushirobira, M.~Korotina, A.~Vedyakov, On preserving-excitation properties of kreisselmeier’s regressor extension scheme, IEEE Transactions on Automatic Control 68~(2) (2023) 1296--1302.
\newblock \href {https://doi.org/10.1109/TAC.2022.3172175} {\path{doi:10.1109/TAC.2022.3172175}}.

\bibitem{Korotina2024}
M.~Korotina, S.~Aranovskiy, R.~Ushirobira, D.~Efimov, J.~Wang, A note on fixed- and discrete-time estimation via the drem method, IEEE Transactions on Automatic Control 69~(7) (2024) 4793--4797.
\newblock \href {https://doi.org/10.1109/TAC.2024.3355803} {\path{doi:10.1109/TAC.2024.3355803}}.

\end{thebibliography}





\end{document}